\documentstyle[epsfig,a4,12pt]{article}
\epsfverbosetrue

\def\q1{\rm{Q_1}}
\def\q2{\rm{Q_2}}
\def\qtot{\rm{Q_{tot}}}
\def\th232{\rm{ ^{232} Th }}
\def\u238{\rm{ ^{238} U }}
\def\cs137{\rm{^{137} Cs }}
\def\ba133{\rm{^{133} Ba }}

\begin{document}
\hfill AS-TEXONO/01-03 \\
\hspace*{1cm} \hfill June 29, 2001

\begin{center}
\Large
\bf{
Studies of Prototype CsI(Tl) Crystal Scintillators\\
for Low-Energy
Neutrino Experiments\\
}
\vspace*{0.5cm}
\large
Y. Liu$^a$,
C.P. Chen$^b$,
H.B. Li$^{b,c}$,
C.H. Tang$^c$,
C.Y. Chang$^d$,
L. Hou$^e$,\\
W.P. Lai$^{b,f}$,
J. Li$^a$,
S.T. Lin$^b$,
C.S. Luo$^b$,
J.F. Qiu$^a$,
H.Y. Sheng$^{a,b}$,\\
C.C. Wang$^c$,
M.Z. Wang$^c$,
S.C. Wang$^b$,
H.T. Wong$^{b,}$\footnote{Corresponding~author:
Email:~htwong@phys.sinica.edu.tw;
Tel:+886-2-2789-9682;
FAX:+886-2-2788-9828.},
B. Xin$^e$,\\
Q. Yue$^a$,
D.X. Zhao$^{a,b}$,
S.Q. Zhao$^{a,g}$,
Z.Y. Zhou$^e$,
B.A. Zhuang$^{a,b}$\\[2ex]
{\large
The TEXONO\footnote{Taiwan EXperiment On NeutrinO} Collaboration
}
\end{center}

\normalsize

\begin{flushleft}
$^a$ Institute of High Energy Physics, Beijing, China.\\
$^b$ Institute of Physics, Academia Sinica, Taipei, Taiwan.\\
$^c$ Department of Physics, National Taiwan University, 
Taipei, Taiwan.\\
$^d$ Department of Physics, University of Maryland, College Park, U.S.A.\\
$^e$ Department of Nuclear Physics, Institute of Atomic Energy, 
Beijing, China.\\
$^f$ Department of Management Information Systems,
Chung Kuo Institute of Technology, \\
\hspace*{1cm} Taipei, Taiwan.\\
$^g$ Unique New Materials Company, Beijing, China.
\end{flushleft}

\clearpage

\begin{center}
{\bf
Abstract
}
\end{center}

Crystal scintillators provide potential merits
for the pursuit of low-energy 
low-background experiments.
A CsI(Tl) scintillating crystal detector
is being constructed to study low-energy neutrino
physics at a nuclear reactor, while
projects are underway to adopt this
technique for dark matter searches.
The choice of the geometrical 
parameters of the crystal modules, as well
as the optimization of the read-out scheme,
are the results of an R\&D program.
Crystals with 40~cm in length were
developed. 
The detector requirements 
and the achieved performance of the prototypes 
are presented. Future prospects for
this technique are discussed.

\vspace*{0.1cm}

\begin{flushleft}
\end{flushleft}

\begin{center}
{\it Nucl. Instrum. Methods. A 482, 125 (2002)}
\end{center}

\clearpage

\section{Introduction}

Crystal scintillators have been widely used
as radiation detectors and in medical imaging~\cite{crystal}.
Large detector systems, of scale several tens of tons in mass,
have been constructed and made operational for
high energy physics experiments~\cite{emcalo}.

The merits of this detector technology
in non-accelerator particle physics experiments
have been recently discussed~\cite{prospects}.
Primarily, the high-Z composition of most crystals allows
a compact design and provides large suppression of
background due to ambient radioactivity if
a three dimensional fiducial volume definition
can be realized.
An experiment based on 100~kg of NaI(Tl) has been
built for Dark Matter searches, producing some
of the most sensitive results~\cite{dama}.

This article reports on the results of
an R\&D program on CsI(Tl) crystal
scintillator~\cite{liuyan} performed by the
TEXONO Collaboration towards
a 500~kg CsI(Tl) detector system to be installed
close to a nuclear reactor core of the
Kuo-Sheng (KS) Power Plant in Taiwan
to study low energy
neutrino interactions~\cite{program,expt}. 
Scintillating CsI(Tl) crystals~\cite{csichar} 
have the merits
of providing large light yield, low energy
threshold and with pulse shape
discrimination characteristics for $\gamma$/$\alpha$
separation. Its property of being only
slightly hygroscopic implies that it
can be machined easily and does not
require hermetic seal (that is, passive materials)
in a large detector system.
In addition, large (40~tons) electromagnetic
calorimeter systems~\cite{bfactories}
have been constructed and 
made operational in high energy physics experiments,
making this technology affordable and
realistic to scale up.
Considering all the associated costs, the price
of CsI(Tl) is in fact less than that for NaI(Tl).

\begin{figure}
\centerline{
\epsfig{file=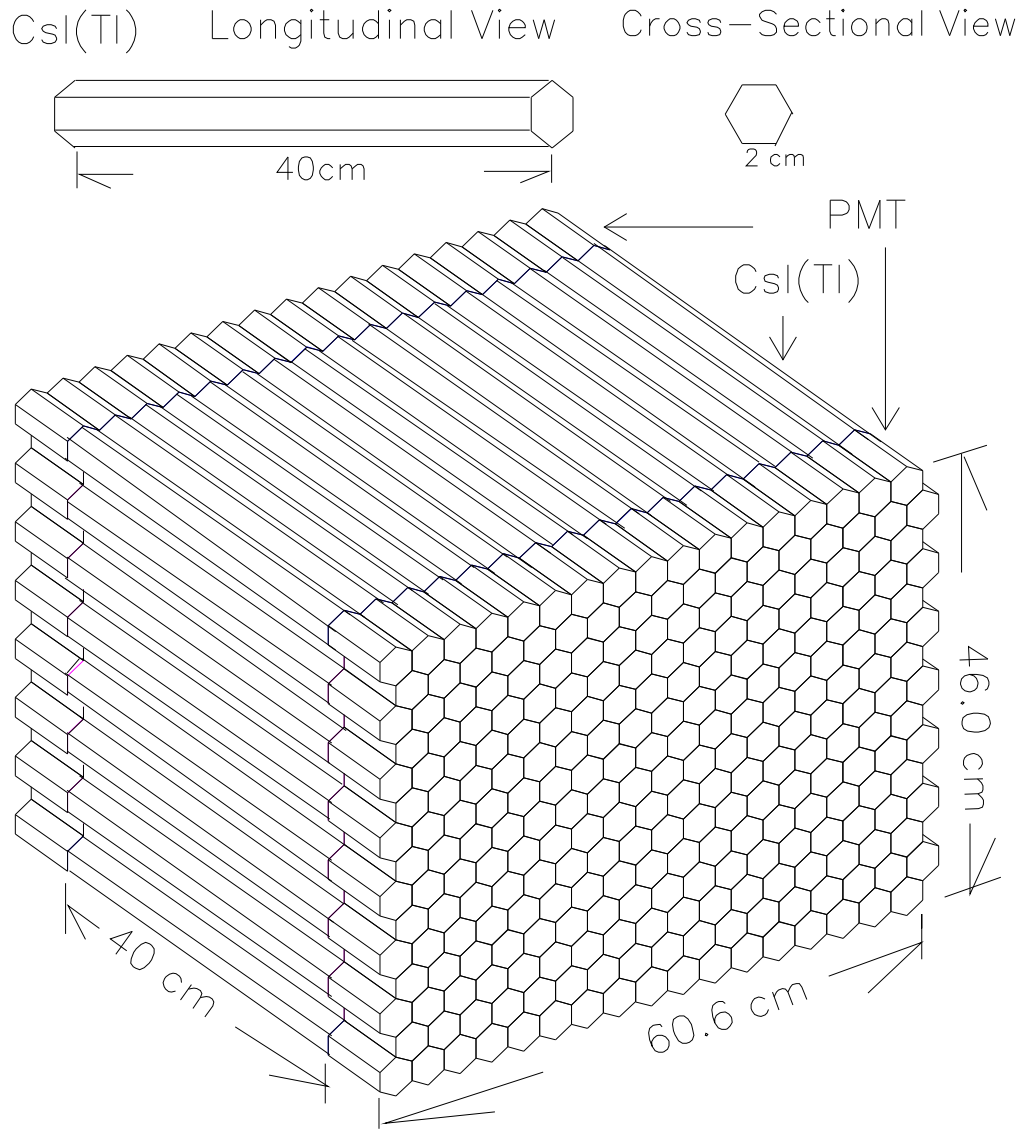,width=15cm}
}
\caption{
Schematic drawings of the CsI(Tl)
detector under construction for the Kuo-Sheng
experiment, showing a 
2(Width)$\times$17(Depth)$\times$15(Height) matrix.
Individual crystal module is 40~cm long with
a hexagonal cross-section of 2~cm edge. Readout
is performed by photo-multipliers
at both ends.
}
\label{csitarget}
\end{figure}

To achieve a three-dimensional fiducial volume
definition with minimal passive materials and dead
space, the detector modules are hexagonal
in shape and stacked together as depicted in
Figure~\ref{csitarget}. 
The system is placed in a dry nitrogen atmosphere
to minimize humidity and to purge the radioactive
radon gas.
The scintillation light
is read out by photo-multipliers (PMTs) at both
ends of the modules. The sum of the signals gives
the energy of the event while their difference
provides information of the longitudinal position.
The PMT signals are shaped and amplified and
subsequently digitized by a 20~MHz 
Flash Analog to Digital Convertor (FADC)
system~\cite{electronics}.
Timing-correlation of possible delayed-cascade
events are also recorded for
the purposes of 
background identification and rejection.

The background from ambient radioactivity
or intrinsic radio-purity are of course
crucial to all low-background experiments.  
Their considerations for CsI(Tl) detectors
are discussed in Refs.~\cite{prospects,expt,csibkg}. 
Levels of better
than the $10^{-12}$~g/g level in
concentration for the $^{238}$U and
$^{232}$Th series have been demonstrated,
assuming secular equilibrium.
With the detector hardware giving 
satisfactory performance,
the background issues will be the principal
focus with the actual data taking on-site
and their control with determine
the sensitivities of the experiment.  

The following sections discuss the choice
of parameters leading to the present detector
design as well as 
the performance of the different options
in the prototype studies.

\section{Scintillating CsI(Tl) Crystal}

\subsection{Fabrication}
\label{sect::fabricate}

The CsI(Tl) crystals are
produced by the Bridgman-Stockbarger method~\cite{bs} in a quartz
crucible~\cite{bscsi} 10~cm in diameter and 
50~cm in length\footnote{Producer: Unique Crystal, Beijing}.
The raw materials include 
CsI powder\footnote{Supplier: Chemetall, GmBH, Germany}
and 1500 ppm of the dopant
thallium iodine (TlI) powder, both purified by baking in vacuum.
The crystal growth 
was also performed in vacuum.
The oven temperature is maintained at 500$^o$C
to 800$^o$C along the crystal growing axis 
with the melting point of 621$^o$C at the growing interface.
The growth rate is about 1~mm per hour. 

The Tl concentration at the 
400~ppm to 2000~ppm range can provide
a reasonable scintillation yield. 
The examination on the uniformity of the light output 
along the longitudinal axis
is therefore necessary to ensure the quality of the crystal.
This requirement of uniform Tl concentration 
places constraint upon the length of crystal growth.
Variation of Tl concentration results in difference
in light yield as well as changes in the rise and decay
times of heavily ionizing events like $\alpha$-particles
and nuclear recoil. Timing response to $\gamma$-events
is unaffected~\cite{csichar}.

Crystal modules produced for the  Kuo-Sheng experiment
are of two categories:
Batch I consists of two pieces of 20~cm crystals glued together
to form a 40~cm module (``L20+20'');
Batch II are CsI(Tl) single crystal of 40~cm in length (``L40'').
The actual growth length is 50~cm from which 40~cm 
is cut out and turned into an active detector.
This is the longest single crystal of CsI(Tl) 
reported for industrial production.
The hot forging method which can be used to
produce large-volume multi-crystalline 
NaI(Tl) detector~\cite{hotnai} is not applicable to produce
large CsI(Tl) multi-crystals. Cracks 
readily develop along the crystal interfaces
upon mechanical stresses.

Both the L20+20 and L40 detector modules are hexagonal in shape
of size 2~cm, giving a cross-section area of 
10.4~cm$^2$, and a mass of 1.88~kg per module.
They can be stacked up into a big fiducial volume
with minimal dead space and passive materials.
The goal of the prototype measurements is 
to optimize  the crystal parameters to 
achieve good energy and longitudinal position
resolutions.

\subsection{Crystal Characteristics}

The main characteristic properties of CsI(Tl) 
scintillating crystals~\cite{csichar},
together with a few other common scintillators,
are listed in Table~\ref{scintab}. 

\begin{figure}
\centerline{
\epsfig{file=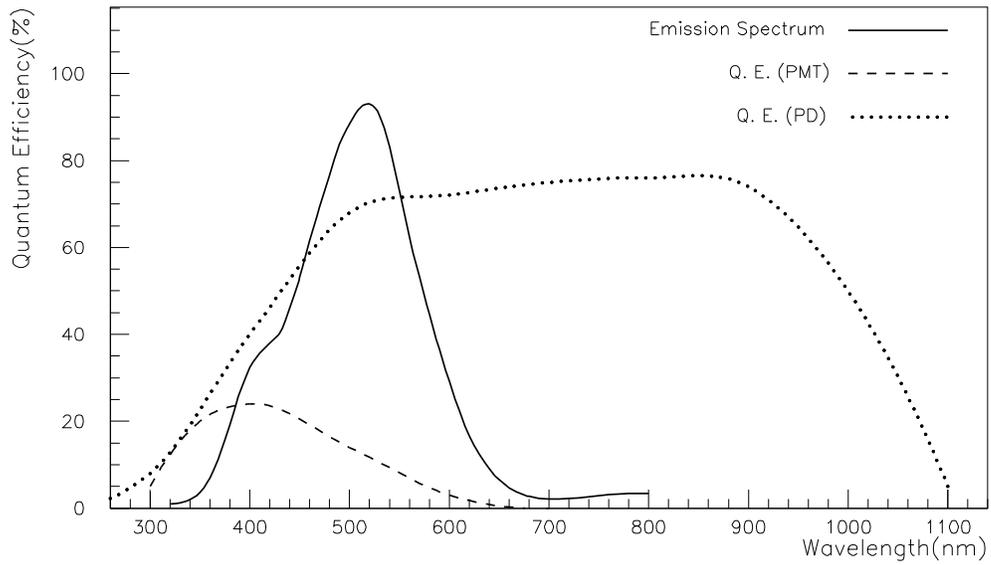,width=15cm}
}
\caption{
Measured emission spectra, in arbitrary units,
from a CsI(Tl) crystal.
The typical quantum efficiencies 
from a photo-multiplier with
bi-alkaline photocathode and from a silicon
photodiode are overlaid.
}
\label{csiemit}
\end{figure}

Undoped CsI emits at ultra-violet (peaks at 310~nm),
has very fast decay time (10~ns), 
and is radiation hard~\cite{csichar,csipure}.
Effects of an admixture of Tl are to absorb the UV emissions,
and shift the spectra to green, via the 
Tl$^+$ luminescence~\cite{scinbasic}.
The emission peaks at about 530-560~nm, as confirmed
by the measured emission spectrum by
a fluorescence spectrophotometer\footnote{Hitachi F-3010},
as displayed in Figure~\ref{csiemit}.
Overlaid on top are typical
response due to  photo-multiplier tube (PMT)
with bi-alkaline photo-cathode and silicon photo-diode (PD).

In principle, CsI(Tl) would produce about 50000 photons
per MeV of electron-equivalence energy deposition,
similar to the range of that for NaI(Tl).
However, since the spectral response of typical PMTs
with bi-alkaline photo-cathode 
do not match well to the emission spectrum, the
effective photo-electron yield is typically
only half of that of NaI(Tl).

\subsection{Pulse Shape Discrimination}

\begin{figure}
{\bf (a)}
\centerline{
\epsfig{file=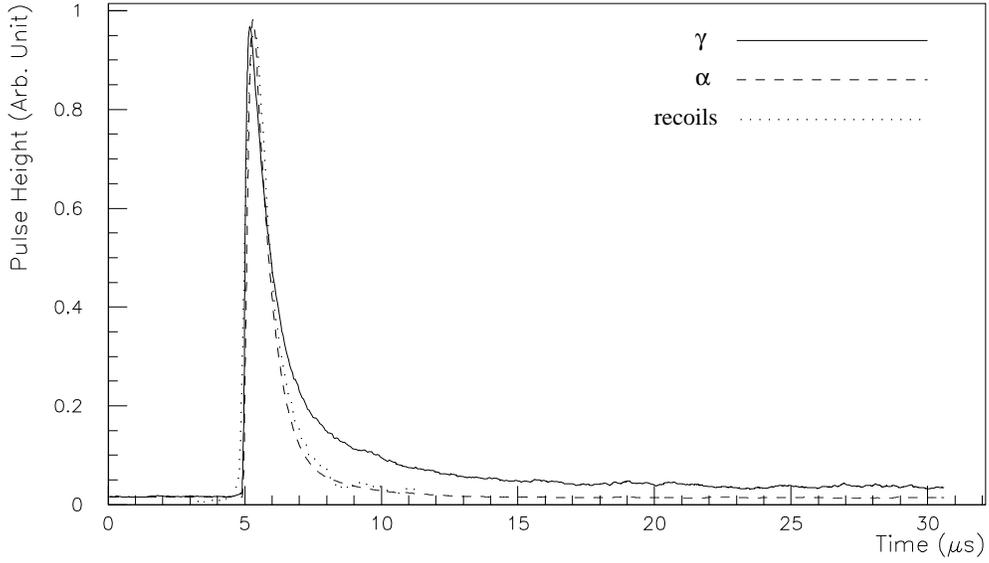,width=15cm}
}
{\bf (b)}
\centerline{
\epsfig{file=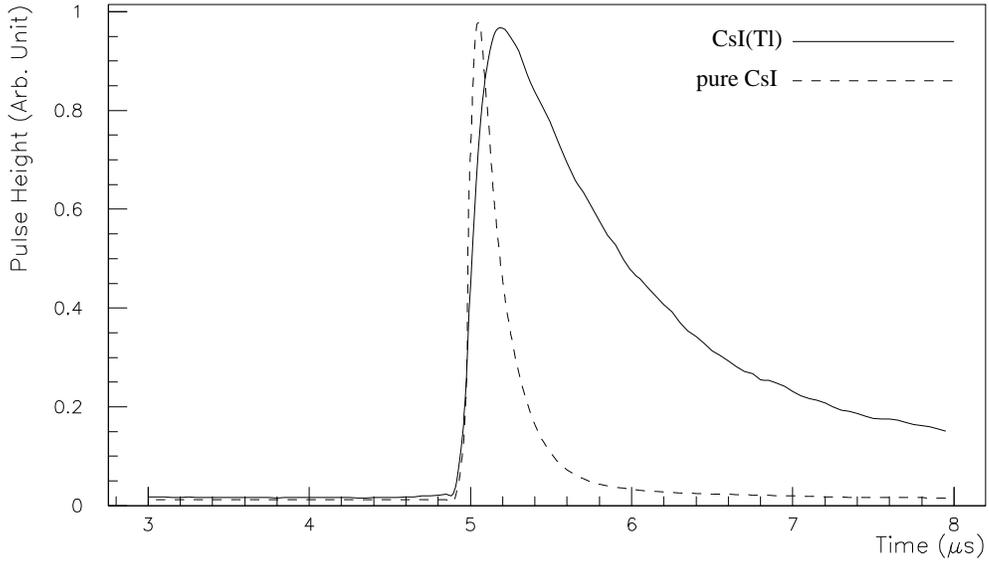,width=15cm}
}
\caption{
The average pulse shape for CsI(Tl)
of (a) events due to $\gamma$-rays at 660~keV, $\alpha$-particles
at 5.4~MeV 
and nuclear recoils at 45~keV,
and (b) its comparison with
an undoped-CsI crystal, as recorded by the
FADC module. The different decay times between
$\gamma$'s and the ions provide
pulse shape discrimination capabilities.
}
\label{psdshape}
\end{figure}

The light emissions  of CsI(Tl)
exhibit different shape for
electrons (that is, minimum ionizing particles)
$\alpha$-particles and nuclear recoils,
as depicted in the
FADC measurement in Figure~\ref{psdshape}a. 
It can be seen that heavily ionizing
events  due to $\alpha$-particles and nuclear
recoils have
{\it faster} decays than those from $\gamma$'s $-$ 
opposite to the response in 
liquid scintillator~\cite{scinbasic}.
This is the basis of the particle identification
of this scintillator~\cite{psdpid}.

Fitted to an analytical form of 
the pulse shape(y) as a function of time(t)
\begin{equation}
\rm{
y ~ =  Constant \ast [ ~ 1 - exp ( - \frac{t}{\tau_0} ) ~  ]
\ast
[ ~ \frac{1}{\tau_1} ~ exp ( - \frac{t}{\tau_1} )  
+ \frac{r}{\tau_2} ~ exp ( - \frac{t}{\tau_2} ) ~ ]
}
\end{equation}
for the light profile from $\gamma$/$\alpha$ events,
one obtains the fitted-values of rise time($\rm{\tau_0}$)
and fall times($\rm{\tau_1 , \tau_2}$)
as well as the ratio between the slow and fast decay
components(r) as tabulated in Table~\ref{time}.
The values of $\rm{\tau_0}$ in CsI(Tl) are dominated by
the electronics shaping rise time
of 250~ns for $> \mu$s pulses~\cite{electronics}.
The intrinsic rise times of the CsI(Tl) 
scintillator are expected to $\sim$125~ns and $\sim$20~ns for
$\gamma$- and $\alpha$-events, respectively~\cite{csichar}.

\begin{figure}
{\bf (a)}
\centerline{
\epsfig{file=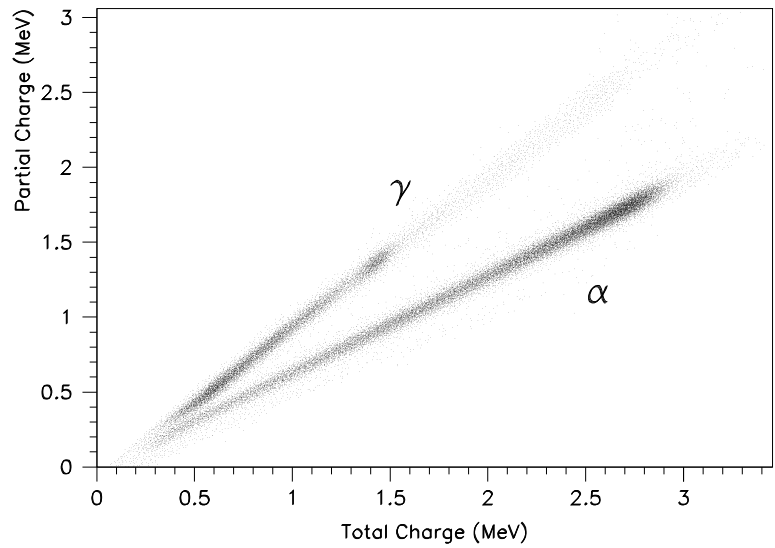,width=12cm}
}
{\bf (b)}
\centerline{
\epsfig{file=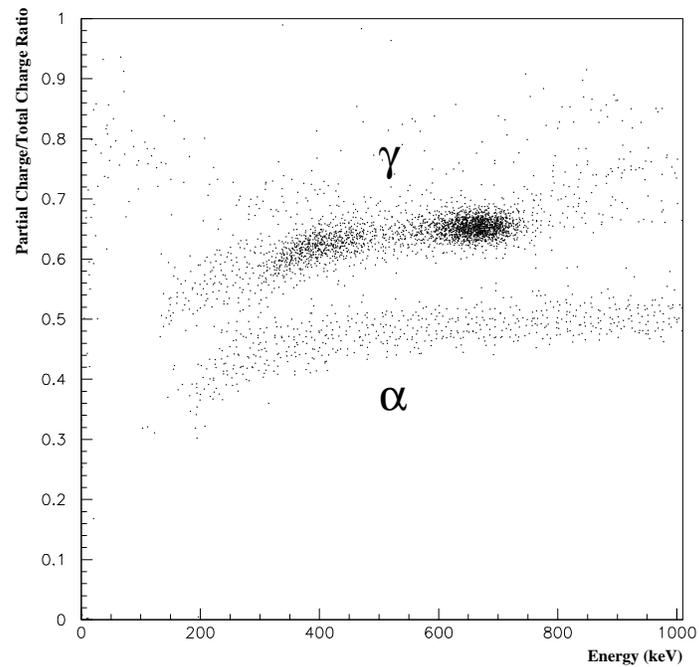,width=10cm}
}
\caption{
The partial charge versus total charge 
at (a) high and (b) low ($<$1~MeV) energy
in a CsI(Tl) crystal,
showing excellent ($>$99\%) pulse shape discrimination
capabilities to differentiate events due to
$\alpha$'s and $\gamma$'s. 
The $\alpha$-events are from 
an $^{241}$Am source placed on the surface of the
crystal. The $\gamma$-events are due to ambient 
radioactivity in (a) and a $^{137}$Cs source in (b).
}
\label{psd2d}
\end{figure}

The difference in the decay time constants between 
$\gamma$ and the ions ($\alpha$ and nuclear recoils)
forms the basis of 
pulse shape discrimination (PSD).
A well-established way to achieve PSD is by
the ``double charge method''~\cite{psddc}.
By comparing the ``total charge'' ($\rm{Q_t}$: integration
over 4~$\mu$s) and the 
``partial charge''($\rm{Q_p}$: integration 
for the same duration {\it after} a delay of 0.5~$\mu$s),
discrimination of $\gamma$/$\alpha$ can be achieved 
with $>$99\% efficiency down to
about 100~keV electron-equivalence light output,
as shown in Figure~\ref{psd2d}.
The difference in pulse shape for nuclear recoils
provides a potential advantage to adopt the 
CsI(Tl) crystal for Dark Matter searches~\cite{csidm}.

Unlike in liquid scintillators, $\alpha$'s are
only slightly quenched in their light output in
CsI(Tl)~\cite{scinbasic}. The quenching factor depends
on the Tl concentration and the $\alpha$-energy~\cite{csichar}.
For full integration of the signals,
the suppression is typically 50\%
at the $\alpha$-energy of 5.4~MeV
for a $^{241}$Am source, as compared to a
typical quenching factor of 10 in liquid
scintillator.  
This small quenching as well as a distinguishable
$\alpha$-signature make this crystal a very
good detector for measuring $\alpha$ and heavy-ions
events in, for instance, the tagging of events
due to internal radioactivity from
the $\u238$ and $\th232$ cascades.

The quenching factors for the
recoils of Cs and I ions are larger, and are typically
at the range of 10$-$20\% at recoil energy less than
50~keV~\cite{csidm,recoil}, 
the energy range relevant for Dark Matter
experiments.

\subsection{Temperature Effects}  

\begin{figure}
\centerline{
\epsfig{file=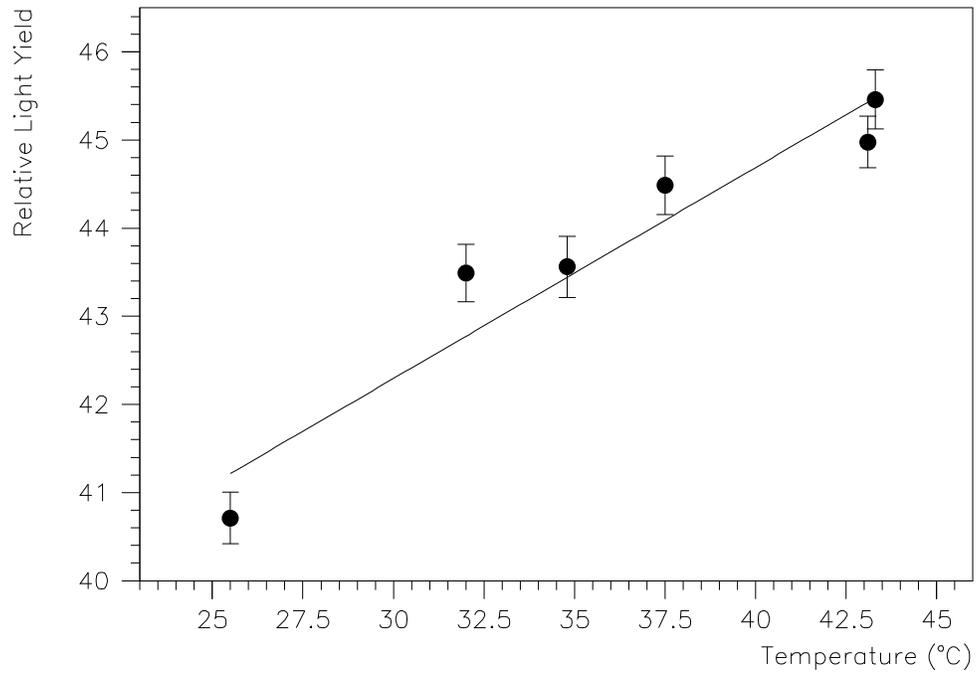,width=15cm}
}
\caption{
Measured temperature effects on the
relative light yield of CsI(Tl)
crystal viewed by a bi-alkaline photo-multiplier
whose temperature coefficients have been
corrected for.
}
\label{temp}
\end{figure}

It is well-established that crystals
like CsI(Tl) and NaI(Tl) give higher light
yield with increasing temperature at typical
laboratory conditions (room temperature).
As depicted in Figure~\ref{temp},
measurements were made to confirm this with
our crystals, giving a enhancement coefficient
of $\rm{ ( 5.8 \pm 0.5 ) \times 10^{-3} ~ ^o C ^{-1} }$.

However, the typical drop in PMT efficiencies
(a typical value
is $\rm{ - 3 \times 10^{-3} ~ ^o C ^{-1}}$ at 525~nm)
as well as the increase of electronic noise in PD readout
with rising temperature
tend to offset this advantage of higher light yield.
Indeed, at the large photo-electrons regime for PD readout
(where performance is limited by electronic noise),
a 10\% better energy resolution was observed by operating
around 5$^o$C where the light yield drops by 10\%.

The results reported in the following sections are
based on measurements done at ``room'' temperature
of about 20$^o$C.

\section{Prototype Detectors}

\subsection{Crystal Geometry}
\label{sect::geom}

To achieve a three-dimensional fiducial volume
definition with minimal passive materials and dead
space, the detector modules are hexagonal (``honey-comb'')
in shape and stacked together as in
Figure~\ref{csitarget}. 
The  crystals are either the L20+20
or the L40 type. 
The length of 40~cm is long compared to
the hexagon cross-section whose sides
are 2~cm. This is to provide a variation
of light collection to establish a 
longitudinal distance measurement.

The surfaces are cut and polished with
aluminium oxide powder to the
``reflection'' (/R) or 
``scattering'' (/S) grade
$-$
the former refers to a surface flatness of better
than 0.5~$\mu$m   which allows total internal
reflection, while the latter has a flatness
of 3~$\mu$m such that light is propagated
by ``diffused'' scatterings. 
The selection between these surfaces is based
on the optimization between energy resolution,
which requires maximum light collection and
position resolution, which requires a certain
level of attenuation effects along the length of the crystal.

\begin{figure}
\centerline{
\epsfig{file=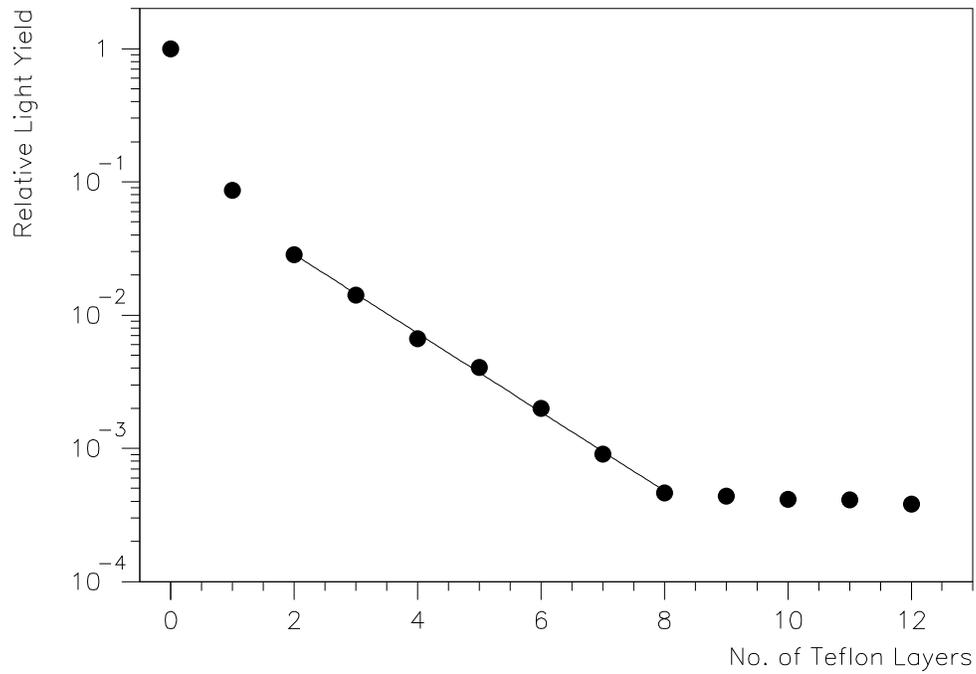,width=15cm}
}
\caption{
The measured pulse height of an LED signal after
traversing a variable number of teflon sheets.
Single photoelectron level is reached after 8 layers.
Error bars are smaller than the symbols.
}
\label{teflon}
\end{figure}

To maximize the light collection, the crystals are wrapped
with four layers of 70~$\mu$m thick teflon sheets
whose purposes are to
provide another diffused scattering surfaces 
to capture the escaped photons back to the crystal,
and to prevent the cross-talks among adjacent
crystal modules. The results to the
cross-talk studies are displayed in Figure~\ref{teflon}
where a LED pulse is illuminated on to a PMT
through a varying number teflon sheets.
It can be seen that the light intensity is 
attenuated by 90\% by a single layer.
Starting from the third layers, the attenuation
factor  is about 
0.5 per layer. These effects can be explained
by multiple-bouncing of the scattered light
within multiple teflon layers.
The total attenuation after 
traversing eight layers (boundary of two adjacent modules)
is $4 \times 10^{-4}$,
ensuring that cross-talks are minimal even for high
energy events up to 10~MeV.

In this intensity setting, the single photo-electron
level was reached after 8 layers where
the ADC-peak positions were not reduced further and
only that the fraction of the ``pedestal'' (sampling
of zeros) events increased.
This gives a convenient
way to monitor single photo-electron responses
of PMTs.

By comparing the photo-peaks of events
due to a $^{137}$Cs $\gamma$-source
for a L20+20 crystal,
the teflon wrappings enhance light collection
by 50\% relative to the one with no
wrappings and collects light only by
total internal reflection from the bare
crystal surface.

\subsection{Readout Device}

In principle, the silicon PIN photo-diode (PD)  gives 
better quantum efficiency  and
matching in the spectral response for
CsI(Tl) crystals than the PMTs with bi-alkali
photo-cathode, as indicated in Figure~\ref{csiemit}.
Indeed, calorimeters in high energy collider experiments
(where the events are of GeV range)
typically adopt PD readout for its compactness and
its being insensitive to magnetic field.

However, the PDs have smaller light collection area
and require pre-amplifiers (PA) for readout. 
The  threshold and energy resolution is dominated
by electronic noise which depends on the dark
current and  input capacitance of the PDs.
Using the 60~keV $\gamma$-ray from $^{241}$Am as
a calibration, one can establish the combined
electronic noise of a 2~cm$\times$1~cm PD\footnote{Hamamatsu S2744-08}
plus PA\footnote{Canberra 2006} is 580~photo-electrons.

\begin{figure}
\centerline{
\epsfig{file=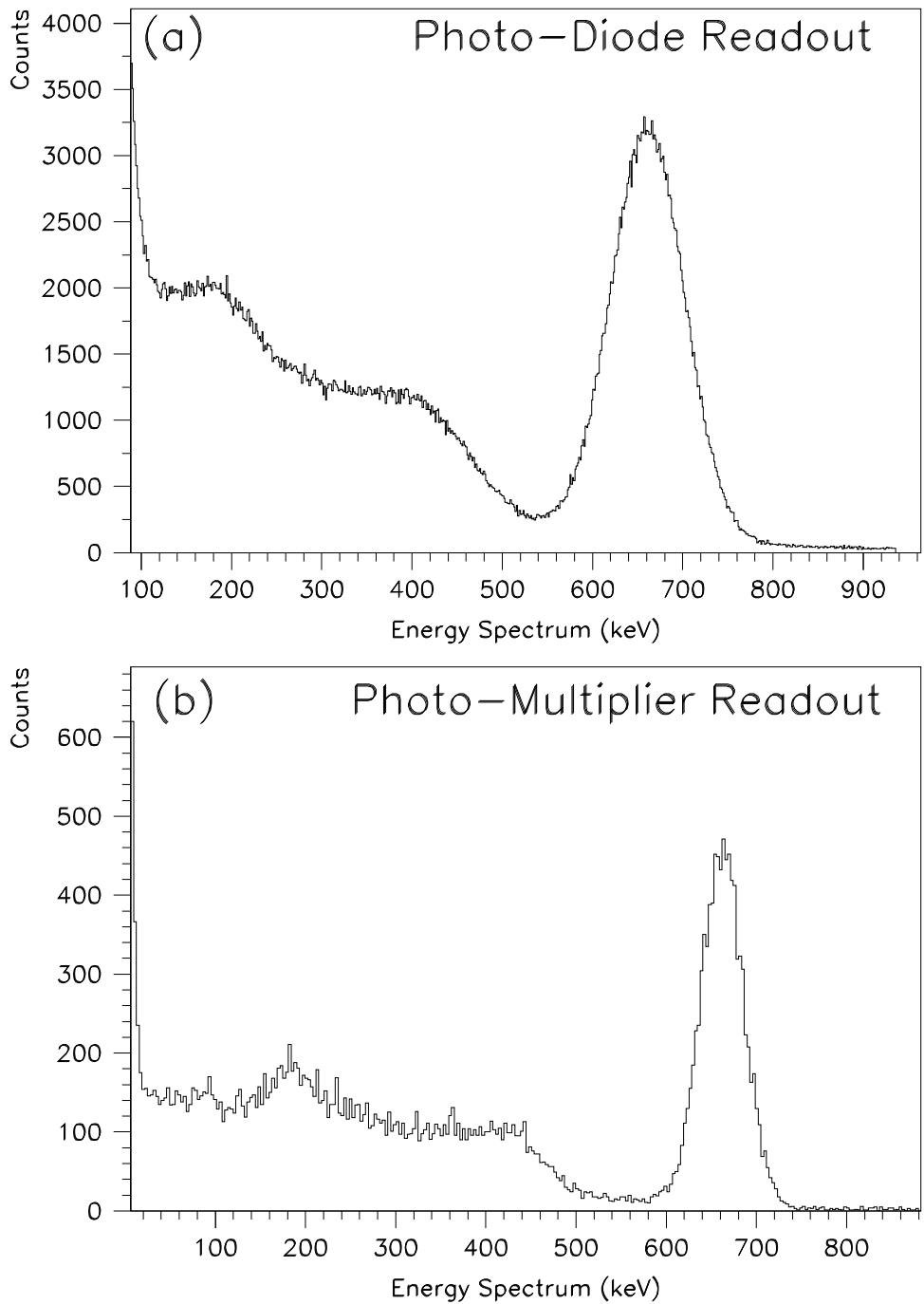,width=15cm}
}
\caption{ Comparison of measured energy spectra
of $^{137}$Cs on a CsI(Tl) crystal
with 2.54~cm diameter and 2.54~cm length 
with (a) photo-diode
and pre-amplifier and (b) photo-multiplier readout.}
\label{pdpmt}
\end{figure}

For low energy experiment on keV-range events,
the PMTs gives superior performance in
terms of resolution and threshold due to
their low noise level of a few photo-electrons,
An illustration of their different
performance can be found in Figure~\ref{pdpmt},
where the measured energy spectra
for $^{137}$Cs by both devices  
on a 2.54~cm diameter and 2.54~cm length
CsI(Tl) crystal are shown.
The FWHM resolution at 660~keV are
14.3\% and 8.4\%, respectively, for PD and PMT. 
The electronic noise threshold also improves from
$\sim$100~keV to $\sim$10~keV from PD to PMT.

Accordingly, a 29~mm diameter PMT\footnote{Hamamatsu CR110 customized}
with low-activity glass window and envelope was chosen
for the experiment. A dynamic range with linear
response of better than
a factor of 1000 is achieved after optimizations
of PMT-base circuit design.
The results reported in subsequent 
sections are based on this PMT
coupled to the crystal surface via optical grease.

\subsection{Prototype Performance}

The performance of prototype modules
(L20+20/S, L40/R, L40/S with definitions
as described in Section~\ref{sect::fabricate})
can be derived from the measurement
with radioactive sources of different
energy. 

\begin{figure}
{\bf (a)}
\centerline{
\epsfig{file=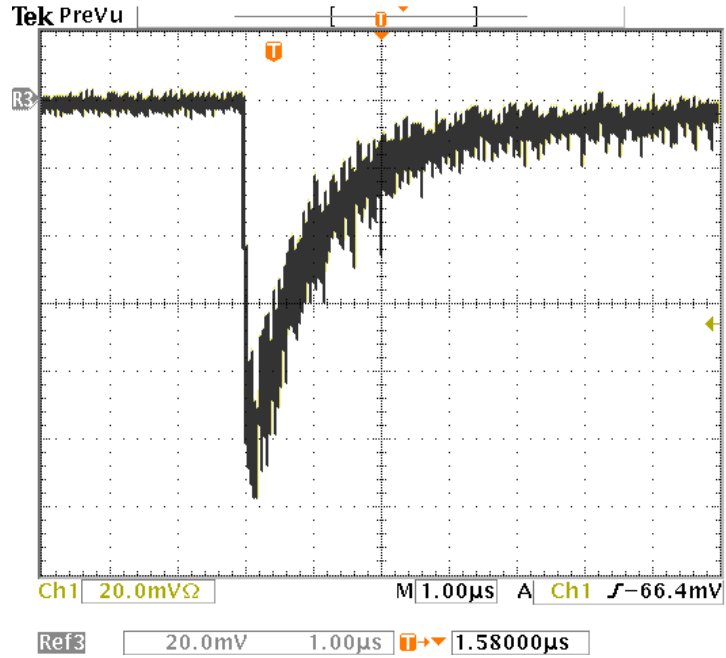,width=12cm}
}
{\bf (b)}
\centerline{
\epsfig{file=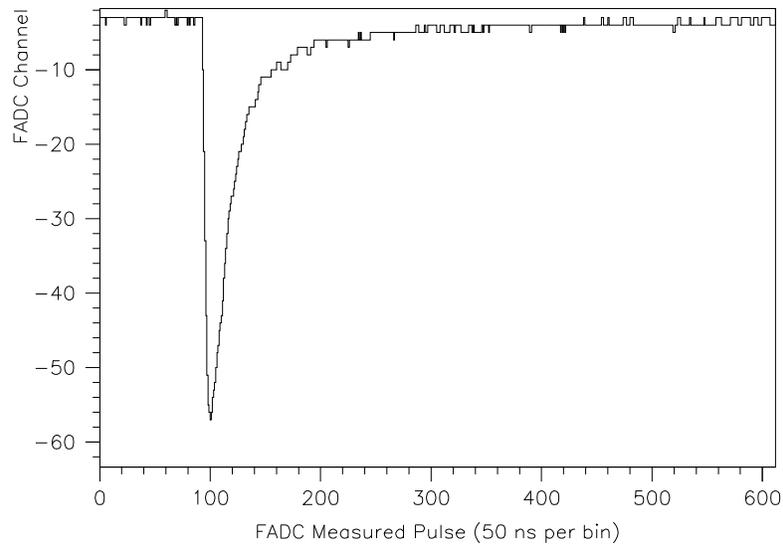,width=12cm}
}
\caption{ 
(a) Raw input signal from CsI(Tl)+PMT as recorded by
a 100~MHz digital oscilloscope.
Time axis: 1~$\mu$s per division.
(b) Output signal
after shaping from the Amplifier-Shaper as recorded
by the FADC.
Time axis: 5~$\mu$s per 100~FADC time bin.
}
\label{singleevent}
\end{figure}

As discussed in Ref.~\cite{electronics},
the  PMT signals are fed into a
Amplifier+Shaper module, the output
of which are recorded by a 20~MHz, 8-bit
Flash Analog Digital Convertor (FADC)
on a VME-bus.
A typical signal due to a $^{137}$Cs source
as measured by the CsI(Tl)+PMT with
a 100~MHz digital oscilloscope
is displayed in Figure~\ref{singleevent}a.
The pulse after electronics shaping and
recorded by the FADC is shown in Figure~\ref{singleevent}b.

The sum of the two PMT signals from
the two ends
($\rm{Q_{tot}=Q_1+Q_2}$) gives the
energy of the event. 
The longitudinal position can be derived 
by the variation of the dimensionless ratio
$\rm { R = (  Q_1 - Q_2 ) / (  Q_1 + Q_2 ) }$
with position.

\begin{figure}
\centerline{
\epsfig{file=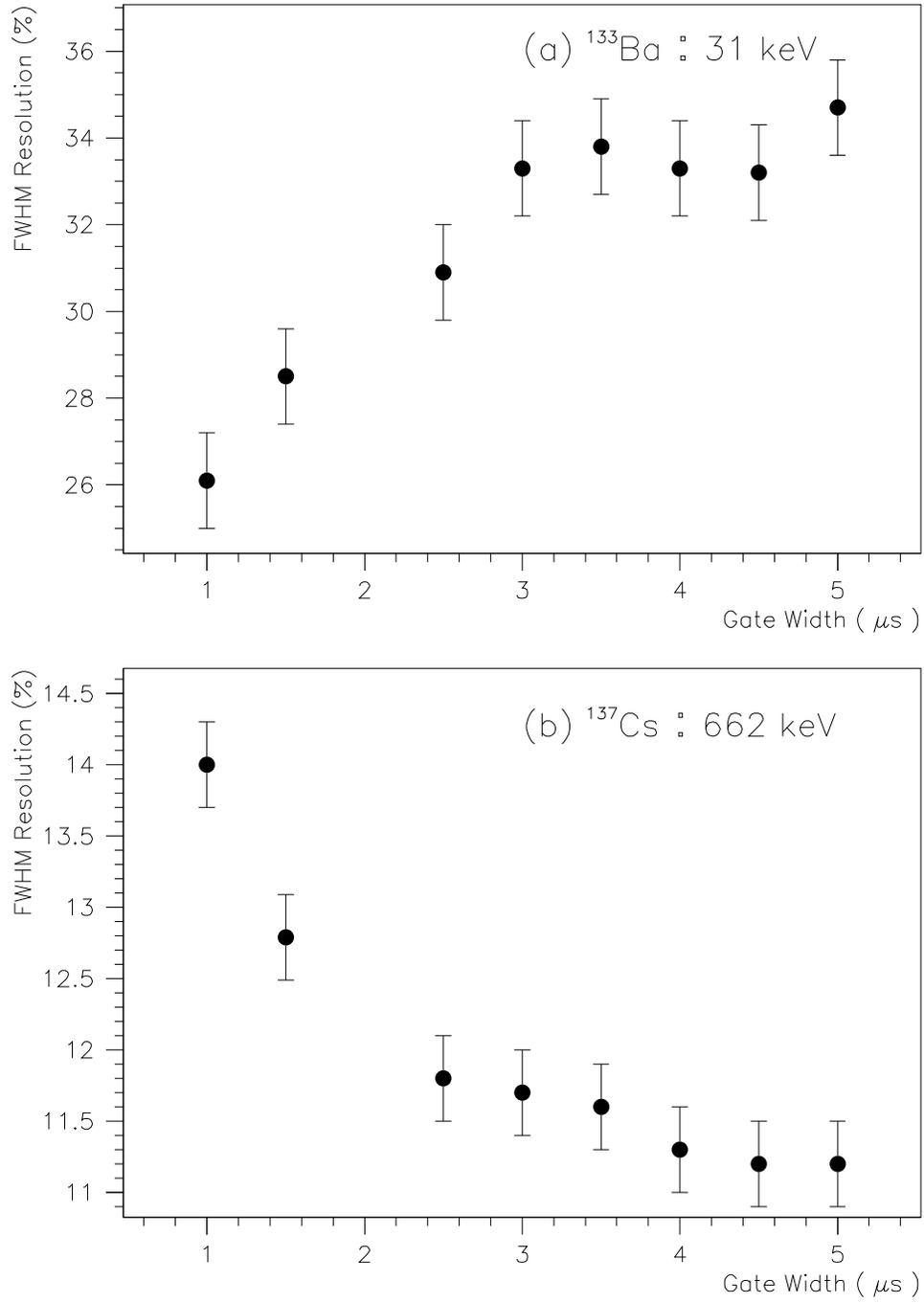,width=15cm}
}
\caption{
Measured FWHM energy resolution
with different integration gate width
for 
(a)  $\rm{E_{\gamma}}$=30~keV for $^{133}$Ba, and  
(b) $\rm{E_{\gamma}}$=662~keV for $^{137}$Cs.
}
\label{gatewidth}
\end{figure}

\begin{figure}
\centerline{
\epsfig{file=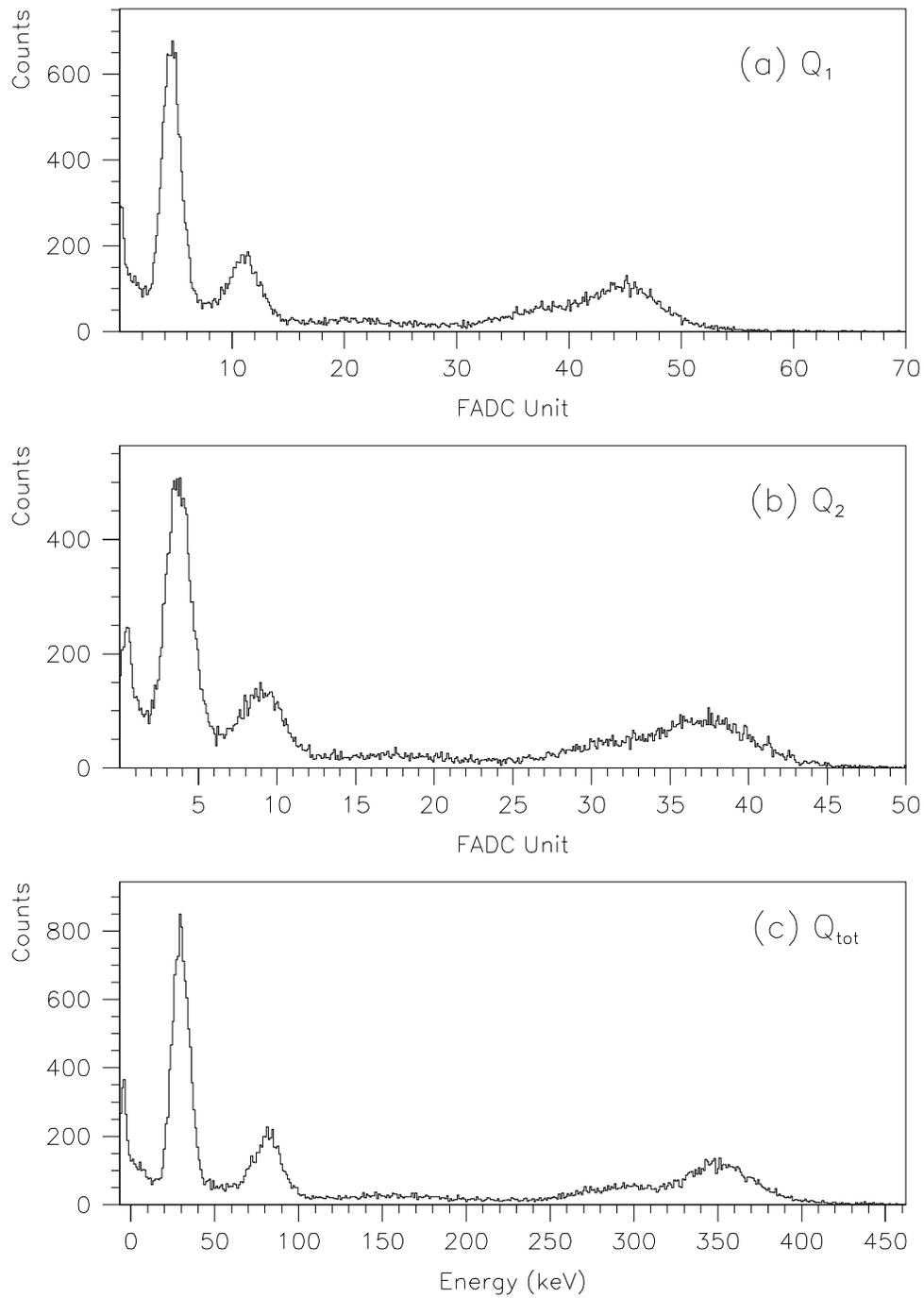,width=15cm}
}
\caption{
Measured spectra with a $\ba133$ source for
(a) $\rm{Q_1}$, (b) $\q2$ and (c) $\qtot$
with the L40/R crystal.
}
\label{ba133all}
\end{figure}

The values of $\qtot$ depends on integration
time defined by software, 
the optimal range of which are different
for different energies. The FWHM resolution of 660~keV
from $\cs137$ and 30~keV from $\ba133$ as function
of integration time are shown in Figure~\ref{gatewidth}
to illustrate this effect. 
Typical spectra of $\rm{Q_1}$, $\q2$ and $\qtot$ due
to a $\ba133$ source for the L40/R crystal
are displayed in
Figure~\ref{ba133all}. It can be seen that
a threshold level of $\sim$10~keV are readily
achievable. The calibration  for $\qtot$ 
is about 1~keV per photoelectron, as derived from
the PMT single photo-electron response with
an attenuated LED source, as discussed in
Section~\ref{sect::geom}.

\subsubsection{The L20+20/S Crystal}  

The L20+20/S module are constructed from
two pieces of 20~cm CsI(Tl) crystals glued
together. Each piece has a light yield uniformity
of better than 5\%  along the 20~cm length,
based on measurements with a PMT coupled to one side 
with the
other end also covered by teflon sheets.

\begin{figure}
{\bf (a)}
\centerline{
\epsfig{file=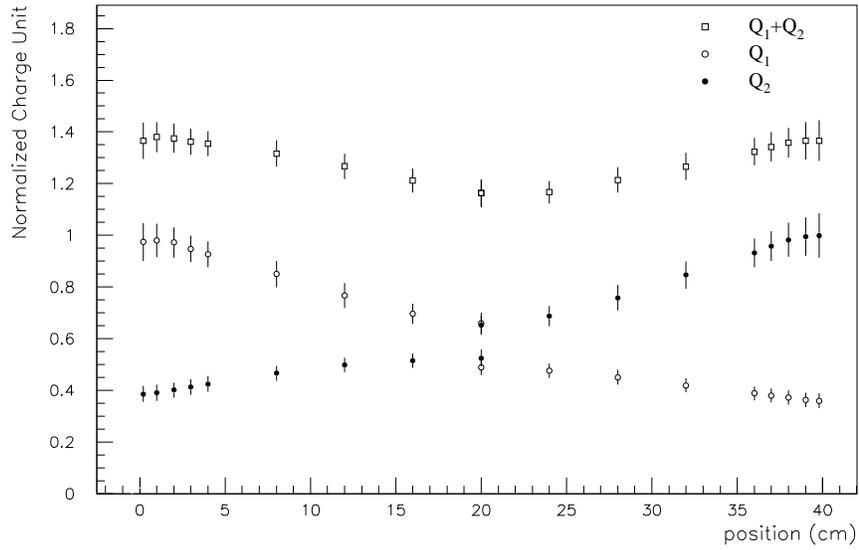,width=13cm}
}
{\bf (b)}
\centerline{
\epsfig{file=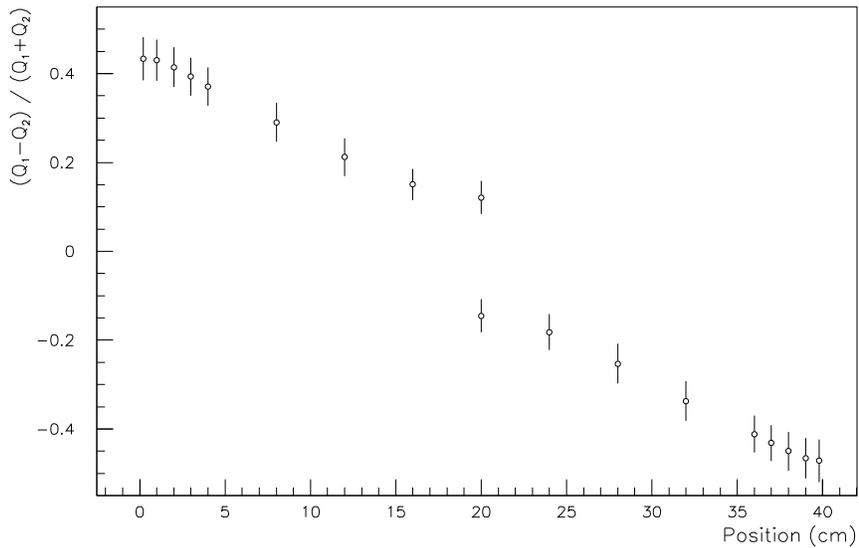,width=13cm}
}
\caption{
(a) Q$_1$, Q$_2$ and $\rm{Q_{tot}= Q_1 + Q_2}$ 
and (b) 
$\rm { R = (  Q_1 - Q_2 ) / (  Q_1 + Q_2 ) }$
along
the longitudinal position of the L20+20/S
crystal module.
The charge unit is normalized to unity for both
Q$_1$ and Q$_2$ at their respective ends.
The error bars in (a) denote the width  of the
photo-peaks due to a $^{137}$Cs source.
}
\label{L20}
\end{figure}

The variation of light
collection
across the length of the crystal
for $\rm{Q_1}$, $\q2$ and $\qtot$ are displayed in 
Figures~\ref{L20}a.
The charge unit
is normalized to unity at the
$^{137}$Cs photo-peak (660~keV) for both
Q$_1$ and Q$_2$ at their respective ends, while
the error bars denote the FWHM width at that energy.
A FWHM resolution of 10\% is achieved. 

The discontinuity at L=20~cm
is due to the optical mis-match between the
glue (n=1.5) and the CsI(Tl) crystal (n=1.8).
Typically, there is a 20\% drop in the
light collection across the glued boundary.
It can be seen that there is a dependence
of $\rm{Q_{tot}}$ with position at the 10-20\% level.

Readout with PMTs at both ends would produce
a larger non-uniformity in total light collection
with position, as compared to light collection
only at one end with the other end covered with teflon
sheets. The teflon serves as a reflecting
surface to bounce back the light which
would otherwise leak out in the case
of the two-end readout configuration.

The longitudinal position can be obtained
by considering the R-value, where its
variations at the $^{137}$Cs photo-peak energy
along the crystal length is displayed in Figure~\ref{L20}b.
The ratio of the RMS errors in R relative to the slope
gives the longitudinal position resolution of
better than 2~cm RMS at 660~keV.

\subsubsection{The L40 Crystal}

The L40 crystals are single crystals of CsI(Tl)
with 40~cm in length.
These were developed and turned into large-scale production 
during the course of our R\&D program. Light
yield uniformity is better than 8\%, while
the decay time profiles are constant to better than 4\%.
In the prototype measurements, crystals with both reflection (/R)
and scattering (/S) grades surfaces were studied.

\begin{figure}
\centerline{
\epsfig{file=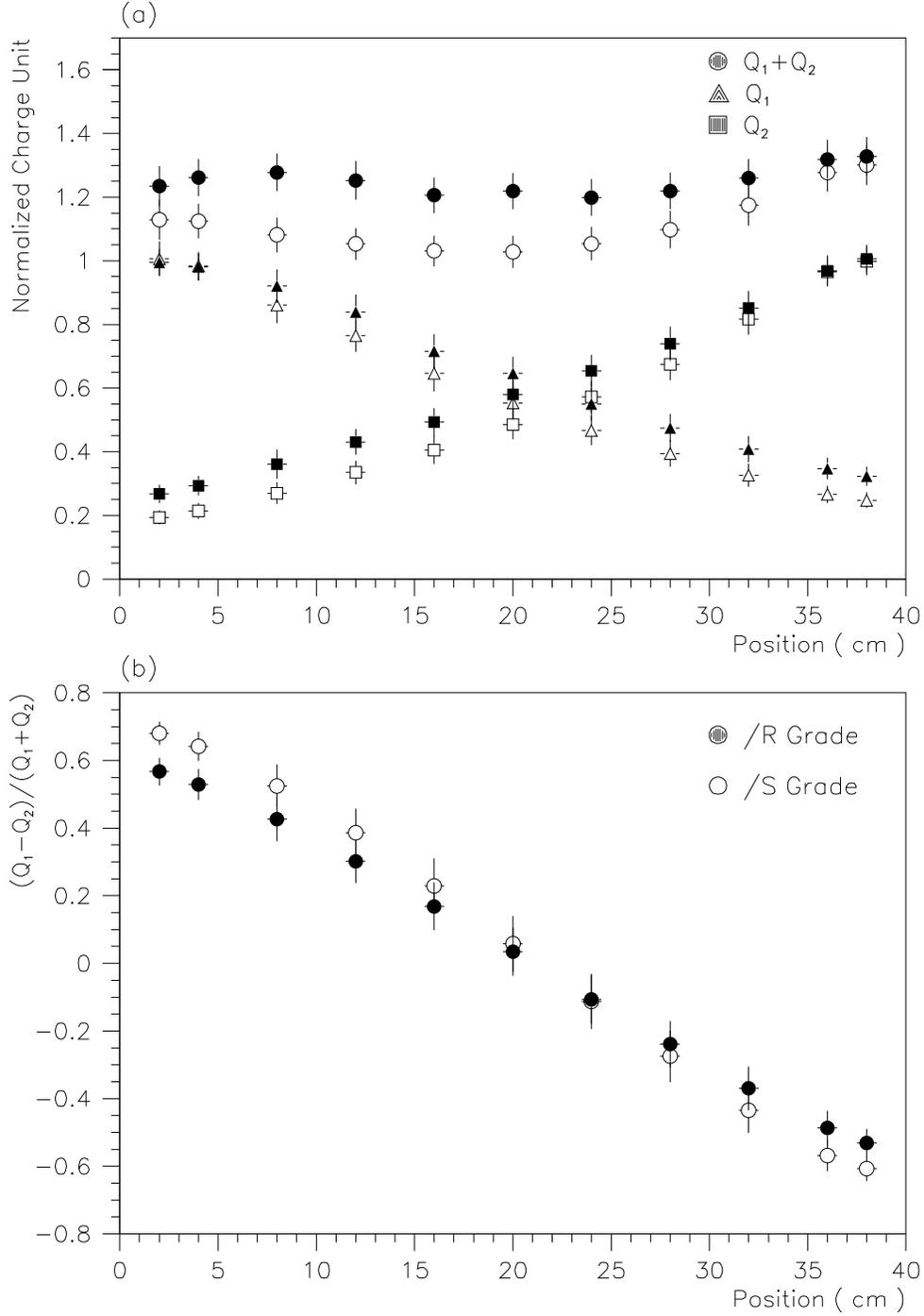,width=15cm}
}
\caption{
The measured variations of
(a) Q$_1$, Q$_2$ and $\rm{Q_{tot}}$
and (b) the R-value
along
the longitudinal position of the L40/R and L40/S
crystal module, denoted by shaded and open symbols,
respectively.
}
\label{L40}
\end{figure}

The light collection and R-value variations across
the length for the L40/R and L40/S
crystals with a $\cs137$ $\gamma$-source
are shown in  Figures~\ref{L40}a and \ref{L40}b, respectively.
It can be seen that 
the light transmission of the /S-grade is slightly
worse than the /R-grade, as indicated by
an ``effective attenuation length'' of 24~cm versus
30~cm, respectively.
The FWHM energy resolutions are 11\%
for both at the central position.
The larger slope in R for the
/S-grade crystal 
is offset by bigger uncertainties, such that
the RMS longitudinal position resolutions
are $\sim$2~cm in both cases.

Therefore, the L40 crystals
give similar performance to the L20+20 crystals
but without the discontinuity 
which complicates the subsequent event re-construction.
The /R and /S grades do not give rise
to big difference in performance.

The variations of
the energy resolution with
energy for both crystal types
are shown in Figures~\ref{L40scan}a.
A FWHM energy resolution of 10\% is achieved  at
660~keV. 
The variations of longitudinal position
resolution with energy
are depicted in Figure~\ref{L40scan}b.
The measured RMS-resolutions
are 3.8~cm  and 6.7~cm at
80~keV and 30~keV, respectively.
Only upper limits of 2~cm can be  derived above 
350~keV, since the interaction points are
no longer localized due to finite
collimator width as well as the dominant 
multiple Compton
scatterings at these higher energy.
There is no energy
dependence of the slopes R to better than 5\%.
The linearity of the energy response
is illustrated in Figure~\ref{L40scan}c,
providing a convenient calibration
using standard $\gamma$-sources.

\subsubsection{Light Guide with Undoped-CsI} 

As indicated in Figure~\ref{L40scan}b, a longitudinal position 
information (or fiducial volume cut)
can be derived for energies above
300~keV. For low energies such as that at the 10~keV
range relevant for Dark Matter experiments, 
the RMS resolution is big such that
the fiducial volume definition becomes inefficient.

An approach to achieve a three-dimensional
volume definition at low energies
is to use active light guide at
both ends.  
The entire fiducial volume
are enclosed by a high-Z, high-density active 
veto volume, suppressing external background.
The light guide materials should have a 
different decay curve from the target
crystal such that events originating at
the light guide can be easily identified 
by pulse shape discrimination.

The undoped-CsI crystal~\cite{csipure} is optimal
for this purpose, having fast decay time ($\sim$10~ns) 
as compared to CsI(Tl) ($\sim$1000~ns)
as depicted by the measured light curves
in Figure~\ref{psdshape}b. 
The fitted values of $\rm{\tau_0}$ and $\rm{\tau_1}$
listed in Table~\ref{time} only reflect the
electronics shaping time 
of 27~ns for fast ($\sim$10~ns) pulses
as well as the finite time bin (50~ns) for the FADC.
The emission spectrum of pure CsI
peaks at the ultra-violet range
and the relative photo-electron yield is only about 0.1 of that
for CsI(Tl) with a bi-alkaline PMT.

\begin{figure}
\centerline{
\epsfig{file=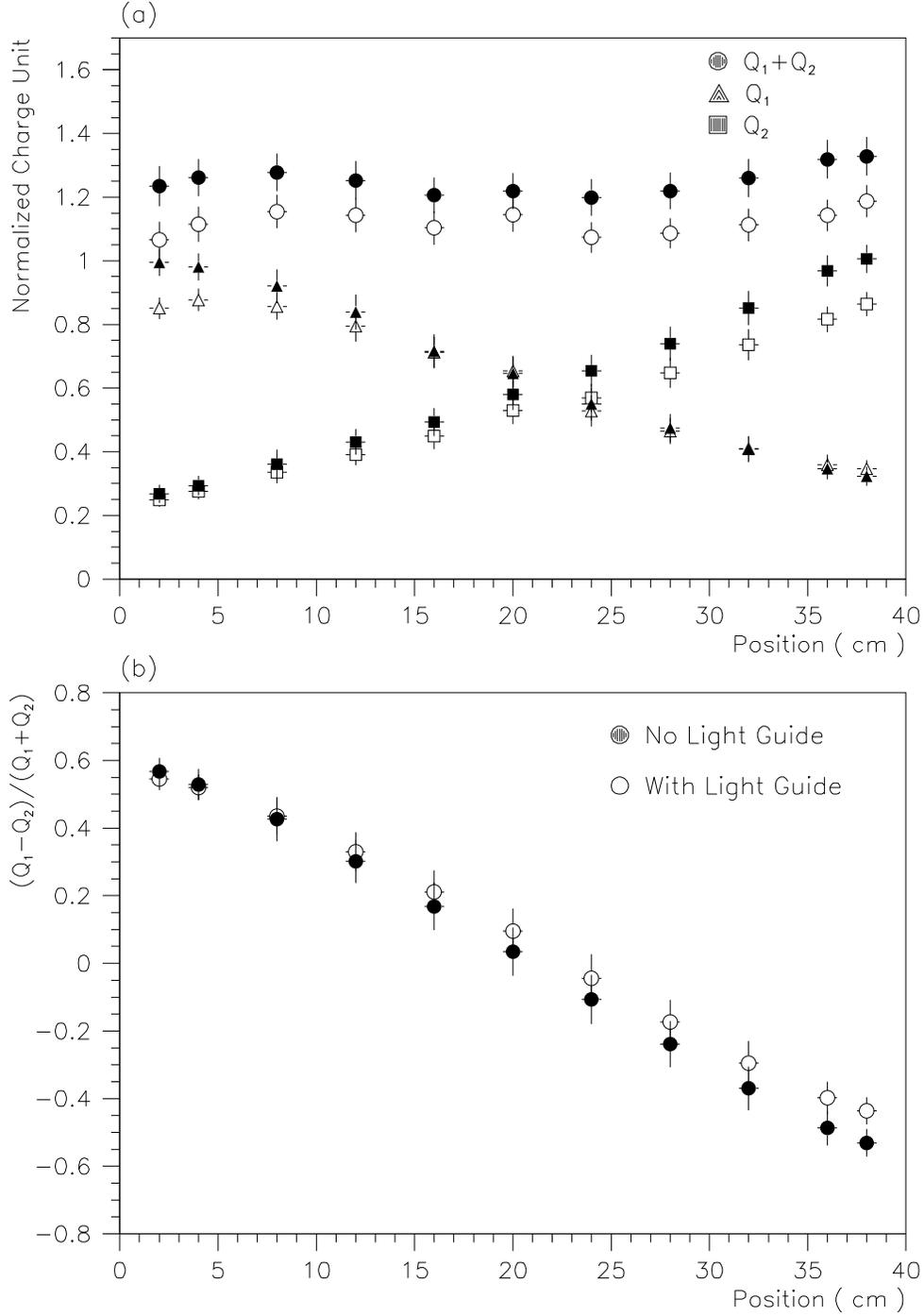,width=15cm}
}
\caption{
Comparisons of L40/R and L40/R+LG crystals,
denoted by shaded and open symbols, respectively,
for (a) Q$_1$, Q$_2$ and $\rm{Q_{tot}}$, and
(b) the R-values. 
The light guides are undoped-CsI with 3~cm in length
at both ends.
}
\label{L40lg}
\end{figure}

Measurements were carried out with the L40/R crystal
with undoped-CsI light guides 
optically coupled at both ends (L40/R+LG).
The light guides have the same hexagonal
cross-section area and are 3~cm in length.
The comparisons of light collection and of the
R-values are displayed in Figures~\ref{L40lg}a and
\ref{L40lg}b, respectively.
It can be seen that
the L40/R+LG module gives similar performance in terms of 
energy and
longitudinal position resolutions,
as compared to the L40/R crystal.

The light collection is inevitably reduced
with the addition of light guides.
It is, however, interesting to note that
the reduction factor is larger ($\sim$10\%) 
at the near end. Only a marginal effect is
observed at the far end. This can be explained
by the fact that light propagation is more forward
(that is, less back-scattering due to 
total internal reflection at the
crystal-glue interface)
for events taking place at the far ends.

The fiducial CsI(Tl) target is totally protected
from ambient radioactivity in a detector
configuration shown in Figure~\ref{csitarget}
but with active light guide at both ends. 
The suppression
is especially strong at low energies. For instance,
the attenuation length of an 100~keV $\gamma$-ray in
CsI is 0.12 cm, such that it is essentially impossible
for external $\gamma$-background at this energy 
to penetrate the 3~cm light guides.

\section{Summary and Prospects}

The research program of the TEXONO Collaboration 
evolves around the theme of exploring  the
use of crystal scintillators for low-energy
low-background experiments
for neutrino and astro-particle physics~\cite{prospects,program}.

An extensive R\&D program was carried out to investigate
the use of CsI(Tl) scintillating crystals.
We report on
the results of the prototype studies in this article.
The major detector requirements and challenges
are the simultaneous
optimization of low threshold, large dynamic
range, good energy resolution,
position resolution and pulse shape discrimination.

\begin{figure}
\centerline{
\epsfig{file=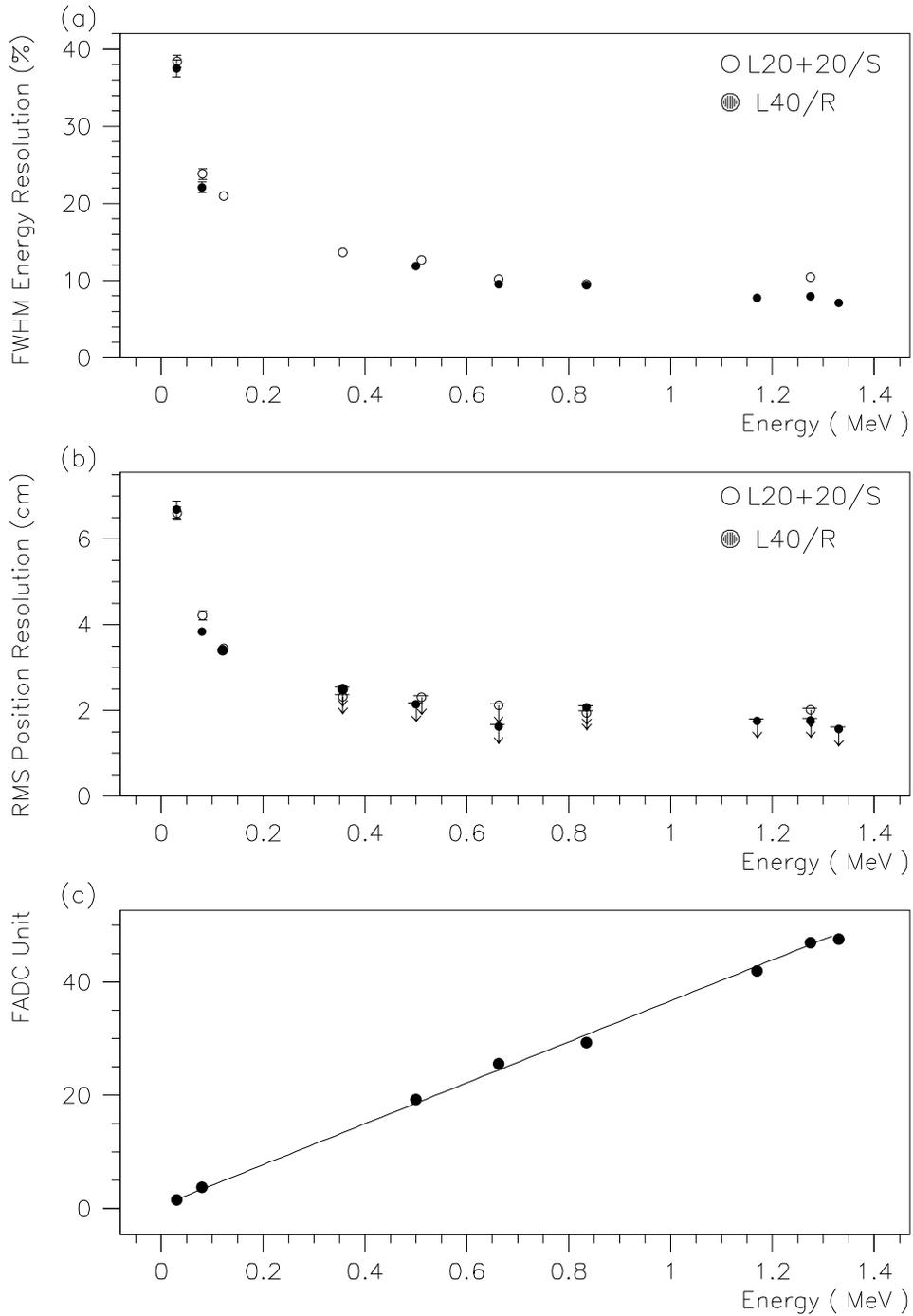,width=15cm}
}
\caption{
The variation of
(a) FWHM energy resolution 
and (b) RMS position resolution
with energy
for L20+20/S and L40/R crystals,
shown in open and closed symbols,
respectively.
Only upper limits are shown
for the higher energy points in (b)
since the events are not localized.
(c) Illustration of the linearity of the
energy response.
}
\label{L40scan}
\end{figure}

Long single crystals of 40~cm  length were successfully
developed and turned into production mode during
the course of the R\&D program.
The performance are similar among the L20+20 and L40 modules,
and so between /S- and /R-grades surface treatment.
The addition of undoped-CsI light guides do not
give notice-able degradation. 
The results displayed
in Figure~\ref{L40scan} 
indicate the capabilities for all these cases,
which have the same modular mass of about 2~kg.
The typical ranges are 10\% for the FWHM energy resolution
at 660~keV, better than 2~cm RMS position resolution at
energy above 400~keV, 10~keV energy threshold, and better
than 99\% PSD capabilities above 100~keV electron-equivalence
energy.
In contrast, the performance can deteriorate dramatically
if  photo-diode readout is used or if the teflon wrapping
sheets are not applied.

For the Kuo-Sheng Reactor Neutrino Experiment~\cite{expt},
a total of 100~kg of the L20+20/S crystals
were produced before the L40 crystals were developed.
Subsequent production will focus on the L40/R crystals.
As discussed, both configurations give similar
performance nominally, but the event reconstruction
algorithms will be simpler for the L40 crystals.
The options were kept 
for using undoped-CsI as light guide for
future upgrades.
Data taking starts in 2001 with 100~kg of crystal modules.

A good longitudinal position resolution,
and hence a  three-dimensional fiducial volume definition,
is possible for energy above 300~keV under the 
present geometry. 
For lower energies, undoped-CsI as active veto and
light guide can be used to provide this desirable
feature.

The measurements under different configurations
reported here can be used to develop
simulation software to study
the characteristic parameters
for light transmission,
reflection and scattering in CsI(Tl)
crystals~\cite{dmproto}.
Such tools would be valuable in the future design
of experiments requiring 
other detector geometries.

This work demonstrates that
the production of 
CsI(Tl) scintillating crystals with
length up to 50~cm is possible.
In addition, different crystals
can be glued together forming larger modules
while maintaining the
good mechanical strength and
optical transmission.
Therefore, it is possible to have detectors
with individual modules $\sim$1~meter in length.
Lateral dimension depends on the size of the crucible
and the range of 15~cm is realistic. 
To maintain similar or better light transmission
and collection, the ``length-to-diameter'' ratio
of the detector module
should be comparable or more than 
10 (40~cm:4~cm in the present case).
As an illustration,
a $\rm{ 15 \times 15 \times 100 ~cm^3}$ CsI(Tl)
crystal would give a modular mass of about 100~kg,
such that a 50~ton detector can be integrated
from only 500~channels.
It is also possible to have a ``phoswich'' detector 
composed of different scintillator crystal modules
and/or light guides glued together. 
The event locations can
be identified by pulse shape discrimination
techniques.

The use of CsI(Tl) crystals for Dark Matter searches were
recently discussed~\cite{csidm}.
The main detector requirements would be to minimize the 
threshold and to perform
PSD for $\gamma$'s and nuclear recoil events at very 
low electron-equivalence (typically few keV) energy~\cite{dmproto}.
Results reported in this article are relevant to
adapt this crystal for Dark Matter experiments.
Light guides will be essential at this low energy
range to define an inner fiducial volume.

Low energy solar neutrino experiment is another exciting
area where scintillating crystals may have potential
advantages, being one of the few technologies where
a wide range of nuclei can be incorporated into
an active detector. Crystals with indium~\cite{snuin}
and lithium~\cite{lii} have been investigated, while
there are much interest recently on 
gadolinium and ytterbium
based crystals~\cite{lens,gso}.
The challenge is to be sensitive to the signal
rate of the range of 1 event per day per 10~tons 
of target materials such that the requirements
of intrinsic radio-purity are extremely stringent.
While the crystals may be different, the essential
design concept of the KS reactor experiment to
achieve a three-dimensional fiducial volume
definition are applicable.

The authors are grateful to the technical staff of our
institutes for invaluable support.
This work was supported by contracts
NSC~88-2112-M-001-007,
NSC~89-2112-M-001-028 and
NSC~89-2112-M-001-056
from the National Science Council, Taiwan,
19975050 from the
National Science Foundation, China,
as well as
CosPa~89-N-FA01-1-4-2 from the Ministry of Education, Taiwan.

\clearpage

\begin{table}
\begin{center}
\begin{tabular}{|l||c|c|c|c|c|c|}
\hline
& & & & & & \\
Properties & CsI(Tl)$^{\ast}$ & CsI(pure) & NaI(Tl) & BGO & Liquid & Plastic  \\
& & & & & &  \\ \hline \hline
& & & & & & \\
Density (gcm$^{-3}$) & 4.51 & 4.51 & 3.67 & 7.13 & 0.9 & 1.0  \\
& & & & & & \\
Relative Yield~$^{\amalg}$ & 0.45 & 0.05 & 1.00~$^{\dagger}$ 
& 0.15 & 0.4 & 0.35 \\
& & & & & & \\
Radiation Length (cm) & 1.85 & 1.85 & 2.59 & 1.12 & $\sim$45 & $\sim$45 \\
& & & & & & \\
dE/dx for MIP (MeVcm$^{-1}$) & 5.6 & 5.6 & 4.8 & 9.2 & 1.8 & 1.9 \\
& & & & & & \\
Emission Peak (nm) & 565 & 310 & 410 & 480 & 425 & 425 \\
& & & & & & \\
Decay Time (ns) & 1000 & 8 & 230 & 300 & 2 & 2  \\
& & & & & & \\
Refractive index & 1.79 & 1.95 & 1.85 & 2.15 & 1.5 & 1.6  \\
& & & & & & \\
Hygroscopic & slightly & slightly & yes & no & no & no  \\
& & & & & &  \\ \hline
\multicolumn{7}{l}{ }\\
\multicolumn{7}{l}{$^{\ast}$ 
At 1500~ppm dopant concentration
}\\
\multicolumn{7}{l}{$^{\amalg}$ 
Effective photo-electron yield from bi-alkaline photo-cathode.
}\\
\multicolumn{7}{l}{$^{\dagger}$ Typical light yield for NaI(Tl) 
is about 40000 photons per MeV.}
\end{tabular}
\end{center}
\caption{Characteristic properties of the common
crystal scintillators and their comparison with
typical liquid and plastic scintillators.}
\label{scintab}
\end{table}

\begin{table}
\begin{center}
\begin{tabular}{|c||c|c|c|c|c|}
\hline
& & & \multicolumn{2}{|c|}{ } & \\
Crystal & Event Type & Rise Time & 
\multicolumn{2}{|c|}{Decay Time Constant}
& Ratio (r) \\ \cline{4-5}
&  & [$\rm{\tau_0}$ (ns)] & Fast Comp. 
& Slow Comp. & \\ 
& & & 
[$\rm{\tau_1}$ ($\mu$s)] & [$\rm{\tau_2}$ ($\mu$s)] & \\ \hline \hline
& & & & & \\
CsI(Tl) & $\alpha$ & 203$\pm$3 & 0.54$\pm$0.1  & 
2.02$\pm$0.02  & 0.29$\pm$0.02 \\
& & & & & \\
CsI(Tl) & $\gamma$ & 261$\pm$2 & 0.87$\pm$0.1  & 
5.20$\rm$0.04  & 0.61$\pm$0.01 \\ 
& & & & & \\
CsI(pure) & $\gamma$ & $\sim$0.55  & 0.19$\pm$0.01  & $-$  & $-$ \\ 
& & & & & \\ \hline
\end{tabular}
\end{center}
\caption{
Fitted rise and decay time constants as
well as the ratio between slow and fast decay components
for $\alpha$ and $\gamma$ events measured by CsI(Tl)
and undoped CsI.
}
\label{time}
\end{table}

\pagebreak

\clearpage

\clearpage

\clearpage

\clearpage

\clearpage

\clearpage

\clearpage

\clearpage

\clearpage

\clearpage

\clearpage

\clearpage

\clearpage

\end{document}